\documentstyle[aps,prl,psfig]{revtex}

\begin{document}
\twocolumn[\hsize\textwidth\columnwidth\hsize\csname
@twocolumnfalse\endcsname
\preprint{ }

\title{See-saw and Supersymmetry \\or Exact R-parity}

\author{Charanjit S. Aulakh$^{(1)}$, 
Alejandra Melfo$^{(2)}$, Andrija Ra\v{s}in$^{(3,4)}$ and Goran
Senjanovi\'c$^{(3)}$}

\address{$^{(1)}${\it Dept. of Physics, Panjab University, 
Chandigarh, India}}

\address{$^{(2)}${\it CAT, Universidad de Los Andes,
M\'{e}rida, Venezuela}}

\address{$^{(3)}${\it International Center for Theoretical
Physics, 34014 Trieste, Italy }}

\address{$^{(4
)}${\it Dept. of Physics and Astronomy,
University of North Carolina, Chapel Hill, NC 27599, USA}}

\maketitle

\begin{abstract}
We show how the renormalizable see-saw mechanism in the context of 
supersymmetry and spontaneously broken B$-$L symmetry 
implies exact R-parity at all energies.
We argue that supersymmetry  plays an important role in providing
a ``canonical'' form for the see-saw, in particular
  in grand unified theories that solve the doublet-triplet
splitting problem via the Dimopoulos-Wilczek mechanism.

\end{abstract} 
 ]
 
\vspace{0.3cm}

{\it A.  The see-saw mechanism} \hspace{0.5cm} 
 The see-saw mechanism \cite{seesaw} is a natural and simple
 way of understanding the smallness of neutrino mass. If one adds a
 right-handed neutrino $\nu_R$ to the Standard Model (SM),
the renormalizable interactions generate 
 a small neutrino mass
\begin{equation}
m_\nu = {m_D^2 \over m_{\nu_R}}
\label{numass}
\end{equation}
where $m_D$ is the neutrino Dirac mass term and $m_{\nu_R}$ is the
Majorana  mass 
of $\nu_R$.
Since we expect the gauge singlet mass $m_{\nu_R}$ to be large, 
$m_{\nu_R}\gg M_W$,
 (\ref{numass}) tells us that $m_\nu \ll m_D$.
The solar and atmospheric neutrino data strongly suggest small 
neutrino masses below 1 eV \cite{neutrino}. For generic $m_D$ of the order 
of charged lepton or quark masses this points to $m_{\nu_R}$ bigger
than $10^{11}$ GeV or so (and possibly as large as the GUT scale).
We stick to this in the rest of the paper. 

Here we address the issue of the see-saw mechanism in 
the context of supersymmetry. This is an extremely important question for
 at least two reasons. First, a central issue of the supersymmetric
 standard model is the fate of baryon and lepton numbers. The conservation
 of B and L is normally connected to  R-parity, and it would be of
great use to have  a more fundamental, underlying principle settling
this question. 
Second, broken R-parity implies nonvanishing
 neutrino masses\cite{am82,hs84} and hence obscures the see-saw
predictions. It is here that the left-right (LR) symmetry \cite{leftright} 
(or B$-$L gauge invariance) plays an
 important role: it
 implies R-parity in the underlying theory \cite{rparity}. Namely, 
 R-parity can be written as
\begin{equation}
{\rm R} =  (-1)^{3(B-L) + 2s} 
\label{matter}
\end{equation}
This means that we must break B$-$L spontaneously in order to generate the
right-handed neutrino mass. We assume that this happens through the
 renormalizable interactions of right-handed neutrinos with a B$-$L=2 
field (singlet under the SM), 
 in order to avoid the forbidden breaking
 of R-parity at high energies $\sim m_{\nu_R}$. We refer
 to this as the renormalizable see-saw. 

Of course, it still remains to
be proved that in the process of coming down to low energies R-parity
is not broken. We carefully  address this issue below and
  prove the following theorem:

{\em The renormalizable supersymmetric see-saw mechanism in theories with
B$-$L symmetry, local or global, implies an exact R-parity even in the low
energy effective
theory}.

This remarkable statement is a simple extension of the impossibility 
of breaking R-parity spontaneously in the MSSM, and it has important
 phenomenological and cosmological consequences. For example, exact 
R-parity ensures
 the stability of the lightest supersymmetric partner, a natural dark
 matter candidate.

Thus, the see-saw mechanism plays a useful role in the determination
 of the structure of the supersymmetric Standard Model (SSM). 
In turn, as we discuss below, supersymmetry
also helps determine the precise form of the see-saw. Namely, in general
in theories beyond the Standard Model the canonical
form as defined in (\ref{numass})
is not complete. We show that supersymmetry may guarantee 
in some instances the canonical form, as for example
in grand unified theories which solve the hierarchy problem by 
the missing VEV mechanism \cite{dw81}.
\vspace{0.3cm}

{\it B. See-saw and exact R-parity } \hspace{0.5cm} The argument here
 is very simple and is a generalization of our recent work on
 supersymmetric LR theories \cite{susylr}.

Let us assume that the original theory possesses a $U(1)_{B-L}$ symmetry,
and the see-saw is achieved by renormalizable terms only.
 Then $m_{\nu_R}$ must be induced through the VEV $\langle
 \sigma \rangle$, where
 $\sigma$ is a B$-$L= 2 superfield. Anomaly cancellation requires the
 existence of  a B$-$L=-2 superfield $\bar \sigma$. The superpotential is
 then given by

\begin{equation}
W_R = {1 \over 2} f\, N^2\, \sigma + g(\sigma \bar \sigma)
\label{superpot}
\end{equation}
where $N$ is the B$-$L=-1 singlet superfield which contains a right-handed 
neutrino $\nu^c \equiv C \bar\nu_R^T$,
and $g(\sigma \bar \sigma)$ is some function of the B$-$L invariant 
combination $\sigma \bar \sigma$.

By properly choosing $g(\sigma\bar\sigma)$, a  non-vanishing
 $\langle \sigma\rangle
\neq 0$ can be enforced \cite{martin}.

 From 
\begin{equation}
F_N = f\, N\, \sigma = 0
\end{equation}
it is clear that $\langle N \rangle =  \langle  \tilde \nu^c \rangle =0 $
 in our vacuum \cite{km}.  Notice that for this to be valid all that we need is
 that $\sigma $ has a non-vanishing VEV, and thus this important result 
does not depend on the details of the model. In particular, it holds
 true in any supersymmetric LR or SO(10) theory with a renormalizable see-saw.
In short, at the scale $M_R$ R-parity remains unbroken,
 as it should be. 

Now what happens as one descends in energy all the way to $M_W$?
 The see-saw  and supersymmetry guarantee a large mass for
 $\tilde \nu^c$: $m_{\tilde \nu^c} = m_{ \nu^c} =  f M_R$ 
and thus the vacuum $\langle \tilde \nu^c \rangle =0$ is stable against any
 perturbation due to soft supersymmetry breaking 
terms.  However, the
 same is not true of $ \tilde\nu$, since it is massless at the scale
 $M_R$. The question of the fate of R-parity is simply a question as
 to whether $ \langle \tilde \nu \rangle $ vanishes or 
not (of course,  $\langle \tilde \nu \rangle \neq 0$ would trigger a 
VEV for 
 $\tilde \nu_c$ through linear terms, but this effect is negligible). 
To see
 what happens, let us recall first the situation with the minimal 
supersymmetric Standard Model (MSSM).

We define the MSSM as the SSM without R-parity breaking
 terms. Then, a nonvanishing VEV for $ \langle \tilde \nu \rangle $ 
implies 
the existence of a 
``doublet''  \cite{am82} 
Majoron \cite{cmp81}, the 
Goldstone boson associated with the spontaneous breaking of the continuous
 lepton number. Such a Majoron is ruled out experimentally \cite{gr81}.
The point is that $ \langle \tilde \nu \rangle $ must be small. 
Consider the most
 conservative case when $\tilde \nu$ is $\tilde \nu_\tau$.
 Since  
$\nu_\tau$ mixes with gauginos 
through $\langle \tilde \nu_\tau \rangle$,
 one obtains an effective mass
\begin{equation}
m_{\nu_\tau} \simeq {\langle \tilde \nu_\tau \rangle^2}/m_{\lambda}
\label{nutau2}
\end{equation}
where $m_{\lambda}$ is a gaugino mass which should lie
below 1 TeV. From the 
experimental limit on $\tau$ neutrino mass one gets
 an upper limit \cite{upper} 
$\langle \tilde \nu_\tau \rangle \leq 10 \,$ GeV.
 Furthermore, if one believes that the
solar and atmospheric neutrino
puzzles are explained by the usual three neutrinos, one gets a much
better limit $m_{\nu_\tau} \leq 5$ eV \cite{bww98}
implying $\langle \tilde \nu_\tau \rangle \leq 10$ MeV.
Now, the scalar partner $R$ of the (pseudoscalar) Majoron, $J$, has a mass
of the order of $\langle
 \tilde \nu_\tau \rangle$ and thus one would have the forbidden decay $
Z \;  \longrightarrow \; J \; + \; R $. In the MSSM,
 it is simply impossible to break R-parity spontaneously.  It is broken
 explicitly or not at all.

 Strictly speaking, in the 
MSSM there is a much more stringent limit on 
$\langle \tilde \nu_\tau \rangle$ from astrophysical considerations.
Unless $\langle \tilde \nu_\tau \rangle \leq  100$ keV, the Majoron would
be
 produced in stars too copiously and radiate their energy away 
\cite{ggn81,am82}.

 Let us now see what happens in the see-saw case. We have
 shown that  $\langle \tilde \nu^c \rangle = 0$, and unless $\langle
 \tilde \nu \rangle \neq 0$ there will be no breaking of R-parity 
whatsoever. Now, once all the fields with mass $\sim M_R$ are integrated out
 the effective theory is the MSSM with all the effects of
 the large scale suppressed, i.e. MSSM + $O(1/M_R)$ effects,
 as dictated by the  decoupling theorem. One obtains an effective 
operator in the superpotential
\begin{equation} W_{\rm eff} = {(L H)^2 \over M_R}
\end{equation}
 where $L$  and $H$
are the lepton doublet and one of the Higgs superfields,
respectively. 
The soft term  $m_{s} W_{\rm eff}$ generates  a tiny mass for the
  Majoron 
\begin{equation}
m_J^2 \simeq {m_s M_W^2 \over M_R} 
\end{equation}
(recall  that the Majoron is predominantly the imaginary component of the
 $\tilde\nu$ field).  
 Since $m_J \ll M_Z$, we end up  with the same prediction
 of the ruled-out $Z$-decay into $J + R$. Surprisingly enough, much 
as in the MSSM, R-parity remains an exact symmetry at all energies.
As such, the argument is independent as to whether B$-$L is a local or
 a global symmetry.

All of the discussion above applies to realistic theories. As we
 have seen, in order to have a theory of R-parity, one needs to
 assume B$-$L symmetry, and this happens automatically in any theory based
 on LR symmetry. Symmetry breaking leading to renormalizable
 see-saw has been studied extensively in \cite{susylr}.

\vspace{0.3cm}

{\it C.  Supersymmetry and the canonical see-saw form}\hspace{0.5cm}
In 
non-supersymmetric theories with LR symmetry one cannot
 ascribe the  ``canonical'' form given in
 (\ref{numass}) to the see-saw. In these theories $\sigma$
 must be a triplet
 $\Delta_R$ of the SU(2)$_R$ gauge group, and the LR symmetry 
implies  the existence of the SU(2)$_L$ triplet
 $\Delta_L$. A simple analysis shows that it must have a VEV too
\cite{ms81}
\begin{equation}
\langle \Delta_L \rangle = \alpha {M_W^2 \over M_R}
\label{deltavev}
\end{equation}
where $\alpha$ is an unknown ratio of the quartic couplings in the
potential.
This stems from  couplings 
\begin{equation}
\lambda \Delta_L \Phi^2 \Delta_R
\label{danger}
\end{equation}
where $\Phi$ is the Higgs multiplet responsible for the fermionic Dirac
 mass terms. One cannot forbid such a term while preserving the see-saw, 
since it is logarithmically divergent at one loop \cite{ms81},
as shown in diagram (a) of Fig. 1 below.
\begin{figure}
\centerline{\psfig{figure=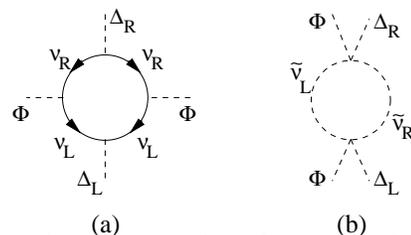,height=3cm}}
\caption{(a) An infinite one-loop  diagram for the interaction $
\Delta_L \Phi^2 \Delta_R$, and (b) its supersymmetric counterpart.}
\end{figure}

Since $\Delta_L$ and $\Delta_R$ are coupled to neutrinos in a LR 
symmetric manner
\begin{equation}
{\cal L}_Y = f ( \ell^T_L C \Delta_L \ell_L + \ell_R^T C \Delta_R \ell_R)
\label{lyuk}
\end{equation}
where $\ell_L$ is the leptonic doublet,
the nonvanishing $\langle \Delta_L \rangle$ provides a direct mass term to 
left-handed neutrinos
\begin{equation}
m_\nu = f \langle \Delta_L \rangle - { m^2_D \over m_{\nu_R}}
\label{typeii}
\end{equation}
 It is still of the see-saw form, {\it i.e.}
proportional to $1/M_R$, but it obscures the 
popular canonical form 
(\ref{numass}) (sometimes referred
 to in the
literature \cite{rabi,cm98}
as type I see-saw, while the see-saw formula with the additional
piece generated by the VEV of $\Delta_L$ 
in (\ref{typeii}) is called type II).

A possible way out of this problem is to break LR symmetry at a much larger
scale than $M_R$ through parity-odd singlets \cite{cmp84} and
suppress $\langle \Delta_L \rangle$.

Another possibility is simply to supersymmetrize the theory. In
supersymmetry no such contribution exists;  diagram (a)
is canceled out by its supersymmetric counterpart,
 diagram (b) in Fig. 1.
 In order to generate  the term in (\ref{danger}) supersymmetry 
must be broken. It is easy to estimate then this  effective term 
which arises from the  difference in right-handed neutrino 
and sneutrino masses.
For the low-energy supersymmetric theory $m_{\tilde\nu_R}^2 \simeq
m_{\nu_R}^2 + M_W^2 = f^2 M_R^2 + M_W^2$ and for $M_R \gg M_W$, we
obtain
\begin{equation}
\lambda \simeq f^2 \sum_g \left({m_D \over M_W }\right)^2 
\ln{ m_{\tilde\nu_R}^2
 \over m_{\nu_R}^2} \simeq \sum_g \left( m_D \over M_R \right)^2
\label{tadpole2}
\end{equation}
where $\sum_g$ is a sum over generations, and for simplicity $f$ is
 taken to be generation-independent. By adding an effective mass term
   $m^2\Delta_L^2$ to the term 
(\ref{danger}), one gets  upon minimization an order of magnitude estimate 
\begin{equation}
\langle \Delta_L \rangle \simeq \left( M_W \over m \right)^2 \sum_g {m_D^2
 \over M_R}
\label{deltavevnew} 
\end{equation}
Notice the factor $( M_W/m)^2$ compared to the canonical see-saw
formula (\ref{numass}). The size of $m$ is model dependent. In
 supersymmetric models based on renormalizable interactions
\cite{susylr} $m$ is of order $M_R$ in which 
case the suppression factor is enormous so that the canonical
form (1) is recovered. In models
based on 
non-renormalizable interactions, though, one has 
instead \cite{susylr} $M_R^2 = m M$, where $M$ is the
 cut-off of the theory. 
For 
$M \simeq M_{Pl}$ one has
\begin{equation}
\langle \Delta_L\rangle = \sum_g {m_D^2
 \over M_R}   \left( M_W^2 M_{Pl}^2 \over M_R^4 \right)
\end{equation}
Only for $M_R$ at its minimum expected value $ \simeq 10^{11}$ GeV
this term can compete with the canonical one, but it becomes rapidly
 negligible with growing $M_R$.

However, this is not the whole story. One can have corrections
to the canonical see-saw even without supersymmetry breaking
from possible
non-renormalizable term in the {\it superpotential} of the form
\begin{equation}
{1 \over M} \Delta_L \Phi^2 \Delta_R^*  \, \, .
\label{wnr}
\end{equation}
Such terms could naturally arise from Planck scale effects
with $M=M_{Pl}$ or, more interestingly, from the GUT scale
physics. 

For example, in the minimal renormalizable supersymmetric SO(10) theory
$M$ is simply $M_X$, the scale of SO(10) breaking. In this model  $M_X
\simeq \langle S \rangle$, where $S$ is the symmetric 54-dimensional
representation, and $\Delta_L$ and $\Delta_R$ belong to a 126-dimensional
representation $\Sigma$. From the superpotential interactions ($\Phi$
is the light Higgs, usually in the
10-dimensional representation)
\begin{equation}
W = \Phi^2 \, S \, + \, \Sigma^2 \, S
\label{so10}
\end{equation}
after integrating out the heavy field $S$, one gets the effective
non-renormalizable interaction (\ref{wnr}), with $M \simeq M_X$.
$\langle S \rangle$ breaks SO(10) down to the Pati-Salam group
SU(2)$_L\times$SU(2)$_R\times$SU(4)$_c$, which is next broken by
a 45-dimensional representation $A$, with $\langle A \rangle =M_{PS}$.
It can also be easily shown that $m \simeq M_{PS}$. 

To find the VEV of $\Delta_L$ we set the F-terms (see \cite{susylr}
for details)
 to zero to get
\begin{equation}
\langle \Delta_L \rangle \approx { {\langle \Phi \rangle^2 
\langle \Delta_R \rangle} \over {mM}}
\equiv\;\epsilon \, {{\langle \Phi \rangle^2} \over 
{\langle \Delta_R \rangle}} 
\end{equation}
where $\epsilon =M_R^2 / {M_{PS} M_X} $.  
 We have  found \cite{amrs99} that
 although a number of new light states 
appear,   succesful unification 
tends to push the intermediate  scales  towards the 
same (GUT) scale,
giving  $\epsilon$ anywhere between $1$ and $10^{-4}$.

The light neutrino mass thus gets the following form:
\begin{equation}
m_\nu \approx ( f^2 \epsilon - h^2_D) {{\langle \Phi \rangle^2}
 \over m_{\nu_R}}
\label{typeiiso10}
\end{equation}
where $f$ is the coupling of $\Delta_L$ to left neutrinos and
$h_D$ is the neutrino Yukawa coupling, and $\epsilon$ is
in general model-dependent. Even for $\epsilon \sim 10^{-4}$ 
the non-canonical part cannot be considered small, 
since for the first two generations the canonical see-saw has
a strong suppression due to the smallness of $m_D$.

 Notice that the undesired operator (\ref{wnr}) originates form the 
exchange of a heavy (3,3,1) field 
(in the SU(2)$_L\times$SU(2)$_R\times$SU(4)$_c$ notation), the
 only field that can couple to both  $\Phi$ and $\Delta_L,\Delta_R$. 
We recover the canonical see-saw  by choosing a  GUT
 scale Higgs that does not contain this field, for example
the 210 representation of SO(10).
 It is noteworthy that 210 contains a parity-odd singlet,
 and thus can give a  canonical see-saw even in 
the non-supersymmetric case (if  $M_R \ll M_X$) \cite{cmp84}.

Another possibility is that even if the GUT scale Higgs contains a
(3,3,1) field, one forbids its  coupling to
$\Delta_R$ and $\Delta_L$ by a discrete symmetry. This is exactly what
happens in the Dimopoulos-Wilczek missing VEV mechanism \cite{dw81}
that solves the doublet-triplet splitting problem. 
In this case, there is
no effective non-renormalizable interaction (\ref{wnr}) (modulo
$1/M_{Pl}$ terms expected to be small for $M_R \leq M_X$). 
An example of an SO(10) model that utilizes the Dimopoulos-Wilczek 
mechanism is given in \cite{bb94} (see also \cite{dw}). In this theory the
Higgs fields are in a pair of ten-dimensional representations $\Phi_1$
and $\Phi_2$. The splitting is achieved with a 45-dimensional 
representation $A$ and the superpotential
\begin{equation}
W = A \Phi_1 \Phi_2 + S \Phi^2_2
\end{equation}
 When  $S$  gets a VEV $\sim M_X$ and $A$
  a VEV ${\rm diag}(a,a,a,0,0)\times\tau_2$, it is obvious
that  both Higgs triplets get heavy, while a doublet Higgs
remains massless. The absence of
the $S \Phi^2_1$ term precisely forbids the
 generation of the troubling non-renormalizable
terms (\ref{wnr}). Thus this
solution to the doublet-triplet splitting
 problem leads to the canonical form for the
see-saw (\ref{numass}). 

Notice that the above  argument could be invalidated if we were to use 
$16$ and $\overline{16}$ instead of 126 in order to generate see-saw
\cite{cm98}. 
However, this choice  
would imply the unacceptable breaking of R-parity at high energies
and the theory would require extra discrete symmetries, contrary to 
the spirit of this paper.

\vspace{0.3cm}

{\it E.  Summary and Outlook}\hspace{0.5cm} The fate of R-parity is 
probably the central issue of
the MSSM. This paper   connects
it to the issue of neutrino mass. We show that {\it the renormalizable
see-saw mechanism through the spontaneous breaking of B$-$L symmetry
implies exact R-parity at all energies}.

On the other hand, in the SSM one could always attribute 
the small neutrino mass to the explicit, albeit small
breaking of R-parity. What we have learned here is that
this is completely orthogonal to the see-saw mechanism:
if the see-saw is operative then simply R-parity never
gets broken.  This should be a welcome result to the
 practitioners of R-parity
breaking, since this mechanism  is then not obscured by the
see-saw as the origin of neutrino mass.

The exact form of the see-saw is model dependent. 
The popular canonical form for see-saw in (\ref{numass}) 
can in general get additional see-saw terms as in
(\ref{typeiiso10}) from higher-scales in the theory.
However, in GUTs that solve the doublet-triplet splitting
problem via Dimopoulos-Wilczek mechanism such extra
terms are absent.

\vspace{0.5cm}

We thank Borut Bajc, Umberto Cotti, and Francesco Vissani for 
 discussions and careful reading of the manuscript. 
We also benefited from discussions with Gia Dvali and Antonio Masiero. 
The work of A.R. and G.S. is partially supported  by EEC 
under the TMR contract ERBFMRX-CT960090 and that of A.M. by 
CDCHT-ULA Project No. C-898-98-05-B. A.R. is also supported by
DOE grant No. DE-FG05-85ER41036. A.M. thanks ICTP for hospitality.


\providecommand{\href}[2]{#2}\begingroup\raggedright\endgroup

\end{document}